\documentclass{emulateapj}

\usepackage{apjfonts}

\shorttitle{Sky Surveys and Quasar Lifetimes}
\shortauthors{Martini \& Schneider}

\slugcomment{ApJL accepted [23 Sept 2003]}

\newcommand{\eg}{{\rm e.g.}}
\newcommand{\etal}{{\rm et al.}}

\newcommand{\tq}{{t_Q}}
\newcommand{\lsst}{{\it LSST}}
\newcommand{\poss}{{POSS}}

\begin{document}

\title{Multiepoch Sky Surveys and the Lifetime of Quasars} 

\author{Paul Martini\altaffilmark{1}}

\affil{Carnegie Observatories, 813 Santa Barbara Street, Pasadena, CA 
91101-1292} 

\altaffiltext{1}{Current address: Harvard-Smithsonian Center for Astrophysics; 
60 Garden Street, MS20; Cambridge, MA 02138; pmartini@cfa.harvard.edu}

\and

\author{Donald P. Schneider}

\affil{Department of Astronomy and Astrophysics, Pennsylvania State University, University Park, PA 16802, dschneider@astro.psu.edu} 

\begin{abstract}

We present a new method to measure the episodic lifetime of quasars with 
current and future large-scale sky surveys. 
Future photometric observations of large samples of confirmed quasars can 
provide a direct measurement (or interesting lower limit) to the 
lifetime of an individual episode of quasar activity ($\tq$) and potentially 
enable the study of post-quasar host galaxies. 
Photometric observations of the quasars found by the Sloan Digital Sky Survey 
(SDSS) and 2dF Survey could, with a time baseline of ten years, determine 
$\tq$ to within a factor of two if $\tq < 10^5$ years, or set a lower limit to 
the quasar lifetime. 
Multiple-epoch, precise photometry with the proposed {\it Large Synoptic 
Survey Telescope} could test more complex models for quasar variability and 
mean quasar luminosity evolution. These observations could also constrain the 
rate that tidal disruptions of single stars produce quasar luminosities. 

It is possible to reverse the order of this investigation; previous-epoch 
plate material, such as the Digitized Sky Survey, can be used to determine if 
any of the SDSS quasars had not yet turned on at the time of these prior 
observations. 
Measurements of the entire SDSS quasar sample over the $\sim 50$ year baseline 
provided by these plates can potentially be used to estimate $\tq$ to within 
a factor of two if $\tq < 10^{5.5}$ years, provided quasar variability can be 
accurately characterized and the detection efficiency and photometric 
calibration of the plate material can be well determined. 
These measurements of $\tq$ will have comparable quality to existing, more 
indirect estimates of the quasar lifetime. 
Analysis of the 3814 quasars in the SDSS Early Data Release finds that $\tq$ 
must be larger than approximately 20,000 years. 

\end{abstract}


\section{Introduction}

The quasar lifetime ($\tq$) is an important timescale for supermassive 
black hole growth and determination of the mechanism(s) that fuel Active 
Galactic Nuclei (AGN). 
The net time that black holes are radiating at quasar luminosities, and 
therefore accreting at approximately the Eddington rate, sets the importance 
of luminous accretion for the production of the present-day space density of 
dormant, supermassive black holes. If this lifetime is longer than 
$\tq > 10^7$ years, then a quasar phase would have led to the accretion of 
much of the mass in the largest present-day supermassive black holes. 
However, the low space density of quasars, even at high-redshift, then implies 
that a quasar phase was relatively rare and not all supermassive black holes 
were quasar hosts. 
In contrast, shorter quasar lifetimes require more of the present-day 
supermassive black hole population to have been quasars, but then to have 
accreted a smaller fraction of their total black hole mass as quasars. 

Most estimates of the net quasar lifetime produce values in the range 
$\tq = 10^6 - 10^8$ years, or lower limits that are consistent with these 
values \citep[for a recent review see][]{martini03}. These estimates are 
generally based on either integral or demographic arguments, which combine 
data on the present-day population of supermassive black holes and the 
accretion by the high-redshift quasar population 
\citep[\eg][]{soltan82,haehnelt98,yu02}, or incorporate quasars 
directly into models for galaxy evolution \citep{kauffmann00}. However, most  
demographic models only constrain the net time that a supermassive 
black hole is radiating above the luminosity threshold for quasars and do not 
address the possibility that a supermassive black hole goes through multiple, 
episodic quasar phases that sum to produce the net lifetime. 
The episodic lifetime is an important parameter for the accretion physics, as 
it is the timescale that an accretion disk can maintain accretion at 
approximately the Eddington rate. 
Existing constraints on the episodic lifetime suggest $\tq > 10^4 - 10^5$ and 
are primarily based on the radiative properties of quasars. These lower limits 
are due to the minimum lifetime required to explain the proximity effect in 
the Lyman$-\alpha$ forest \citep{bajtlik88,scott00} and the sizes of 
ionization-bounded narrow-line regions \citep{bennert02}. A recent observation 
of the transverse proximity effect in He II does imply a longer lifetime of 
$\tq \sim 10^7$ years \citep{jakobsen03}; however there is currently only 
one such observation. 

Measurement of both the episodic and net lifetimes of quasars are important 
to determine the methods that trigger quasar activity, as well as 
the physics of the accretion process. If net and episodic lifetimes are 
similar, and by implication long, quasars could be produced by the 
relatively rare merger of approximately equal-mass, gas-rich galaxies. 
This is currently the most favored mechanism invoked to explain the 
mass accretion rates necessary to fuel quasars, as well as their rapid 
decrease in space density at low redshifts \citep{carlberg90,barnes92}. 
In contrast, if the episodic lifetime of quasars is significantly less than 
the net lifetime, then quasar activity must be triggered by a phenomenon that 
is more frequent than large galaxy mergers. 
Some fraction of quasars could also be powered by the tidal disruption of 
individual stars as they pass near peribarathron \citep{hills75,young77}, which 
could produce quasar luminosities for on order one year \citep{ulmer99}. 
The current estimates and limits on both the 
net and episodic lifetimes are still uncertain by several orders of magnitude
due to the quality with which the current data are fit by models, and 
the assumptions of the models themselves. In the present work we describe a 
method that could improve the limit on the episodic lifetime by approximately 
two orders of magnitude and determine if the episodic and net lifetimes are 
comparable. 

The most direct method to determine $\tq$ is to actually measure a quasar 
turn on and off. This is impractical for any given quasar, given the long 
lifetime implied by the methods described above, but is possible for 
observations of an ensemble of quasars, such as are now being produced 
by the 2dF \citep{croom02} and SDSS \citep{york00} surveys. 
The 2dF survey has produced a final catalog of 23,424 quasars, while the 
SDSS survey plans to complete a catalog of 100,000 quasars over the next 
several years. 
In the next section we provide a detailed calculation of how well $\tq$ can be 
computed from these spectroscopically-identified quasar samples when compared 
with photometry from at least one additional epoch. 
We also describe how the SDSS quasar catalog could be compared to the existing 
{\it Palomar Observatory Sky Survey} (\poss), particularly in light of 
the lower-quality and comparable sensitivity of these plates. 
This method is illustrated through application to the SDSS Early Data Release 
(EDR) quasar catalog.
We conclude with a discussion of the implications of these measurements or 
upper limits on $\tq$ in the final section. 

\section{Method} 

We shall first examine the problem of quasar lifetimes with a simple model 
for the quasar luminosity: the quasar is either ``on'' or ``off'' with a 
characteristic on lifetime $\tq$. In this scenario, the probability that any 
quasar will turn off over a time baseline $\Delta t$ is 
$P_{off} = \Delta t(z)/\tq$, where $\Delta t(z) = \Delta t/(1+z)$ is the 
time baseline in the quasar frame from the spectroscopic identification to the 
second, photometric epoch. For a sample of $N$ quasars where 
only $N_{on}$ are observed to be on at the second epoch, the lifetime 
can then be directly calculated as 
\begin{equation}
\tq = \frac{\Sigma \Delta t(z_i)}{\Sigma i - N_{\rm on} }
\end{equation}
where this assumes that a quasar will definitely be detected if it is on at 
the second epoch. Here we adopt the same definition of a quasar as the 
SDSS EDR quasar catalog: an AGN brighter than $M_{i^*} = -23$ mag 
\citep{schneider02}. 
For a spectroscopically-identified quasar with one additional, photometric 
epoch we consider a quasar as on if it is above this luminosity threshold, 
and off if it is fainter than this limit. The limitations of 
this assumption are discussed in \S\ref{sec:dis}. 

\begin{figure}
\epsscale{1.1} 
\plotone{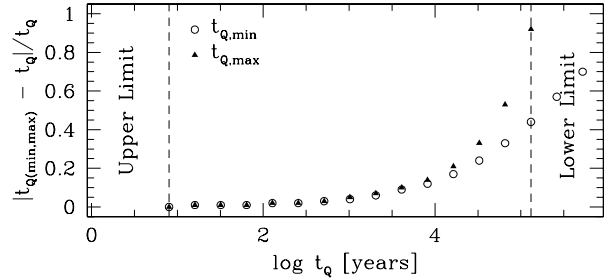} 
\caption{\small
Quality of the potential measurement of the quasar lifetime by the \lsst\ as a 
function of quasar lifetime. The constraint or measurement of $\tq$ is based 
on reobservations of 120,000 quasars ten years after their epoch of 
spectroscopic identification, which provides a net time baseline of 
approximately 500,000 years for the redshift distribution of the SDSS EDR. 
The magnitude of the lower ({\it open circles}) and upper ({\it filled 
triangles}) bounds on $\tq$ have been computed with Poisson statistics. 
\label{fig:lsst}
}
\end{figure}

One survey that could provide the second epoch photometric observation for the 
SDSS and 2dF quasar samples is the proposed {\it Large Synoptic Survey 
Telescope} \citep[\lsst;][]{tyson02}. 
The \lsst\ is intended to regularly survey the entire visible sky to a depth 
many magnitudes fainter than the limit of the SDSS and 2dF quasar samples. This 
telescope will therefore easily obtain high-quality photometry of all of the 
quasars in the final 2dF and SDSS catalogs provided they are still on at the 
time of the \lsst\ observations. 
We have modeled the quality with which the SDSS and 2dF samples could be used 
to measure the quasar lifetime by generating mock catalogs with a range of 
lifetimes and the same magnitude and redshift distributions as the SDSS 
EDR. 
Figure~\ref{fig:lsst} shows the quasar lifetimes and uncertainty in $\tq$ 
attainable with this second epoch \lsst\ observation. 
If the \lsst\ obtains a second baseline measurement of these 120,000 quasars 
ten years after their initial spectroscopic identification, then this will 
measure $\tq$ to within a factor of two if $\tq < 10^5$ years, or otherwise 
set a lower limit to the quasar lifetime if no quasars have been observed to 
turn off. The upper and lower limits to the quasar lifetime are set by the 
length of the time baseline and the probability that no quasars will actually 
be observed to turn off over this baseline, respectively. 
For long values of the lifetime only a small number of quasars will be 
observed to have turned off; we have calculated the one-sided, $1-\sigma$ 
confidence limits on the number of quasars that are off using Poisson 
statistics as described in \citet{gehrels86}. 

For second-epoch survey material that does not extend many magnitudes below 
the limits of the spectroscopic sample, the sensitivity limit, quasar 
variability, and the quality of the photometric calibration could 
all produce a false signal of off quasars. 
These additional effects can be accounted for by assigning a weight $w_i$ to 
each quasar that quantifies the probability that it could be detected 
at the second epoch, given an assumed magnitude distribution based on the 
spectroscopic epoch, a parameterization of quasar variability, and the 
sensitivity of the second epoch photometric survey: 
\begin{equation}
w_i = \int dm_{ij} \; P_m(m_{ij} | \mu_i) \; P_d(m_{ij})
\end{equation}
The expected magnitude distribution for the quasar $P_m(m_{ij} | \mu_i)$ can be 
calculated from a Gaussian distribution centered on its assumed mean magnitude 
$\mu_i$ at the time of identification, as well as the photometric uncertainty 
and quasar variability. 
Due to a variant of the Eddington bias, this mean magnitude will on average be 
fainter than the magnitude $m_i$ at the time of spectroscopic selection 
because of quasar variability $\sigma_V$, the photometric 
uncertainty $\sigma_P$, and the slope of the quasar number-magnitude relation.  
The variability of quasars is commonly parametrized with 
a structure function \citep{simonetti85,hughes92} and long term monitoring of 
quasars shows that the amplitude of the variability increases with time
\citep{hawkins02}. 
The width of the distribution of likely quasar magnitudes at the second epoch 
is thus the quadrature sum of the photometric uncertainty and the 
time-dependent variability:  
$\sigma (\Delta t) = \sqrt{\sigma_P^2 + \sigma_V^2(\Delta t)}$. 
This magnitude distribution is then convolved with the detection probability 
$P_d(m_{ij})$ for a quasar with magnitude $m_{ij}$ at epoch $j$. 
With this weighting function the total number of expected quasars in the 
second epoch becomes the sum of these weights $\Sigma w_i$ and the total time 
baseline available becomes $\Sigma w_i \Delta t(z_i)$. 
The equation for the lifetime is rewritten:  
\begin{equation}
\tq = \frac{\Sigma w_i \Delta t(z_i)}{\Sigma w_i - N_{\rm on} }. 
\end{equation}

A particularly useful set of archival observations are the \poss-I 
plates obtained in the 1950s. 
All of these observations have been digitized, are publicly available on 
CDROM, and provide the longest time baseline for the SDSS quasar catalog. 
As above for the \lsst\ calculation, we have used the redshift and magnitude 
distribution of the SDSS EDR quasars to model the sensitivity of the 
SDSS $+$ \poss-I to the quasar lifetime. 
For regions which were not covered by the \poss-I survey, we have used the 
more recent \poss-II data from the 1980s. 
While the 23,000 2dF quasars will also be valuable when combined with future 
\lsst\ observations, they are not suitable for comparison with the archival 
plate material as the spectroscopic quasar candidates were identified from 
these plates.

\begin{figure}
\epsscale{1.1} 
\plotone{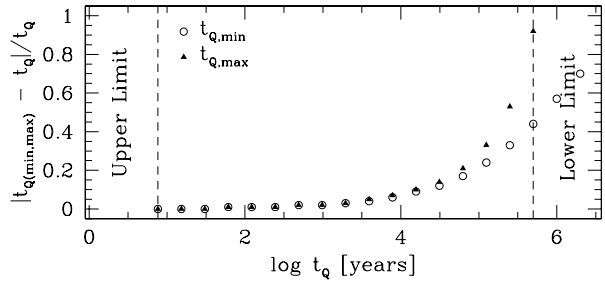} 
\caption{\small
Same as Figure~\ref{fig:lsst} for a comparison of the 100,000 quasars expected 
in the SDSS sample with the \poss\ observations. The time baseline in this case
is approximate two million years. In practice, the constraint on $\tq$ from 
the \poss\ comparison will be limited by uncertainties in quasar variability 
and the \poss\ photometric calibration. 
\label{fig:poss}
}
\end{figure}

The first step in this analysis was the photometric calibration of the \poss\ 
plates, which was carried out with USNO stars selected from the same 
plates as the EDR quasars, to have similar colors \citep{richards02}, and 
small proper motions. 
This produced a catalog of $\sim 9800$ stars that were used to calibrate the 
\poss\ plates directly onto the SDSS magnitude system. The rms uncertainty in 
the 
photometric calibration is $\sim 0.4$ mag. These same stars 
were then used to determine the detection efficiency as a function of 
magnitude by measuring the number of stars identified with 
SExtractor \citep{bertin96} in $0.5$ mag bins. The USNO catalog is complete 
to $V = 21$, which is fainter than the \poss-I limit and the magnitude for 
SDSS quasar selection. The EDR quasars were detected using the same 
SExtractor parameters and requiring that the source on the \poss-I plate 
be within $4''$ of the SDSS coordinates. 
Finally, quasar variability was parameterized with the structure function 
calculated by \citet{hawkins02}. This structure function, which is similar to 
that recently derived from the SDSS EDR quasars by \citet{devries03}, 
was used to set the width of the expected quasar variability distribution 
as a function of time baseline in the quasar frame. 
Figure~\ref{fig:poss} shows the quasar lifetimes and uncertainty in $\tq$ 
attainable with a combination of the 100,000 SDSS quasars and the 
\poss\ epoch. These data will thus in principle be able to measure the quasar 
lifetime if $\tq < 10^{5.5}$, although in practice the sensitivity will be 
somewhat less due to uncertainties in quasar variability and the photometric 
calibration of the plates. Application of this method to the SDSS EDR sample 
sets a lower limit of $\tq > 20,000$ years on the episodic quasar lifetime. 

\section{Discussion} \label{sec:dis}

The present lower limit of $\tq > 20,000$ years from the SDSS EDR quasar 
sample is in agreement with the lower limits to $\tq$ from the radiative 
arguments outlined in the Introduction. 
This limit can be 
improved by approximately the factor of 30 increase in sample size expected 
from the final SDSS quasar sample. The potential measurement or 
limit of $\tq > 10^{5.5}$ years from the full sample will either measure 
$\tq$ or produce an improved lower limit to the episodic lifetime of quasars. 
Future observations of the full quasar sample with the LSST and a ten-year 
baseline could push the lower limit to $\tq \sim 10^6$ years through use of 
the combined baseline provided by \poss\ and \lsst\ data. These observations 
will also be able to set interesting limits on the rate at which tidal 
disruptions of stars produce quasar luminosities. 

The measurement of $\tq$ via \lsst\ observations of the complete SDSS and
2dF quasar samples will provide a comparable constraint on $\tq$ to 
many of the existing, 
indirect estimates from demographic arguments, which typically find values 
of $\tq \sim 10^7$ years \citep{haehnelt98,kauffmann00,yu02}. If the 
sky-survey measurement or lower limit and these other estimates of $\tq$ 
remain comparable in the face of improved data, this will be a strong 
validation of the assumptions of radiative efficiency, black hole mass 
function, and Eddington fraction that are key components of the 
demographic models. 

Observations with the \lsst\ will also provide superb measurements of quasar 
variability, which motivates a more careful consideration of our definition of 
the episodic quasar lifetime. 
A particularly important question is what is meant by a quasar that is 
off, relative to a quasar that is instead experiencing extreme variability. 
Here we have defined a quasar as off if it becomes fainter than the 
absolute magnitude $M = -23$ used to define the quasar sample. 
This definition is motivated by the small number of available epochs for the 
quasars in the \poss\ data and the 
poor photometric precision of the plates, which make it difficult to 
characterize the variability of individual quasars. 
Our definition, however, allows for the spurious identification of off 
quasars due to misclassified, extremely variable objects, or simply variable 
quasars near the luminosity limit that defines the quasar class. 
In principle, a quasar should not be classified as off if it has 
temporarily faded to become a lower-luminosity AGN and then 
subsequently increases in brightness to become a quasar again. 
For quasars with only a small number of epochs, particularly of poor 
photometric precision, this point will be hard to circumvent. 

The multiple-epoch and greater precision observations with the \lsst, however 
will allow this point to be addressed in two ways. First, extremely variable 
quasars can be identified and not included in the sample through a 
variability cut for the lifetime estimate. A second approach is to 
include even these extremely variable quasars, yet require that they 
fade several magnitudes below the defining luminosity of a quasar for some 
number of years before they are formally 
classified as off. This later 
approach is more appealing as eliminating the most variable segment of the 
quasar population may eliminate quasars that are near the end 
of their lives. 
If the quasar has turned off, whether completely or is simply temporarily 
quiescent, then this will offer an unprecedented opportunity to study the 
quasar host galaxy. 

The luminosity and variability evolution of quasars can be considered 
in greater detail with the multiple-epoch \lsst\ observations. 
We have adopted the simple assumption that quasars undergo no luminosity 
evolution while they are on, other than their known variability about some 
constant mean. 
Many models for quasars, however, adopt an exponentially decaying luminosity 
\citep[\eg][]{haehnelt98}. If the mean luminosity of a quasar 
decays exponentially over an $e-$folding timescale $\tq'$, rather than 
experiencing a simple shut off after $\tq$, then the SDSS and \lsst\ photometry 
could be used to search for a mean decrease in the luminosity of the 
quasar population, provided it is relatively short. For an assumed 
photometric uncertainty of $0.02$ mag and a baseline of ten years, 
a dataset of 100,000 quasars could in principle measure a change in 
mean luminosity for a characteristic decay time of close to $\tq' \sim 10^5$ 
years, although in practice quasar variability will limit the sensitivity 
of such a measurement to much shorter characteristic decay timescales.

Longer lifetimes, whether in the simple on--off model or with more complex 
luminosity evolution, could naturally be even better investigated with 
yet larger samples. While a significantly larger spectroscopically-identified 
sample is unlikely in the near future, a larger color-selected sample is 
possible. Based on studies with the color-selection algorithms used for the 
SDSS quasar sample, there are regions of color space that are almost 
exclusively occupied by quasars \citep{richards02}. 
Thus large numbers of quasars could be 
identified through color-selection alone, either in the SDSS or with \lsst,
and then monitored over a number of years. While the remaining contaminating 
population must be considered, these are almost exclusively non-variable 
objects such as white dwarfs, and many of these contaminants could be 
identified via their proper motions. 

Completion of the SDSS should also produce improved estimates of the net 
lifetime of quasars from other techniques such as quasar clustering 
\citep{martini01a,haiman01}. A preliminary comparison of the 2dF clustering 
measurements to the models in \citet{martini01a} suggest the quasar lifetime 
may be as short as $\tq \sim 10^6$ years. 
If refined estimates based on SDSS measurements continue to point toward a 
shorter lifetime, then the net lifetimes determined from clustering will be 
comparable to the range of lifetimes probed by the method outlined here and 
these two methods could determine if the net and episodic lifetimes are 
comparable. If they are, then supermassive black holes only experience one 
quasar phase and the mechanism that ignites and fuels quasar activity is 
a relatively rare event. In contrast, if the episodic lifetime is much shorter 
than the net lifetime, then quasar activity must be triggered by additional 
mechanisms than just the relatively rare mergers of gas-rich galaxies. 

\acknowledgements 
We acknowledge helpful comments from Pat Hall, Gordon Richards, Martin Gaskell, 
and the referee, Elaine Sadler. 
PM was supported by a Carnegie Starr Fellowship and DPS was partially supported 
by National Science Foundation Grant NSF-03007582.
Funding for the creation and distribution of the SDSS Archive
has been provided by the Alfred P. Sloan Foundation, the
Participating Institutions, the National Aeronautics and Space
Administration, the National Science Foundation, the U.S.
Department of Energy, the Japanese Monbukagakusho, and the
Max Planck Society.
The SDSS is managed by the Astrophysical Research Consortium
(ARC) for the Participating Institutions.  The Participating
Institutions are The University of Chicago, Fermilab, the Institute
for Advanced Study, the Japan Participation Group, The Johns
Hopkins University, Los Alamos National Laboratory, the
Max-Planck-Institute for Astronomy (MPIA), the
Max-Planck-Institute for Astrophysics (MPA), New Mexico
State University, University of Pittsburgh, Princeton University,
the United States Naval Observatory, and the University of
Washington.
The Digitized Sky Surveys were produced at the Space Telescope Science
Institute under U.S. Government grant NAG W-2166.
The images of these surveys are based on photographic data obtained using the
Oschin Schmidt Telescope on Palomar Mountain and the UK Schmidt Telescope.
The National Geographic Society - Palomar Observatory Sky Atlas (POSS-I) was
made by the California Institute of Technology with grants from the National
Geographic Society.
The Second Palomar Observatory Sky Survey (POSS-II) was made by the California
Institute of Technology with funds from the National Science Foundation, the
National Geographic Society, the Sloan Foundation, the Samuel Oschin
Foundation, and the Eastman Kodak Corporation.


\begin{thebibliography}{}
\expandafter\ifx\csname natexlab\endcsname\relax\def\natexlab#1{#1}\fi

\bibitem[{{Bajtlik} {et~al.}(1988){Bajtlik}, {Duncan}, \&
  {Ostriker}}]{bajtlik88}
{Bajtlik}, S., {Duncan}, R.~C., \& {Ostriker}, J.~P. 1988, \apj, 327, 570 

\bibitem[Barnes \& Hernquist(1992)]{barnes92}
Barnes, J.E. \& Hernquist, L. 1992, \araa, 30, 705

\bibitem[{{Bennert} {et~al.}(2002){Bennert}, {Falcke}, {Schulz}, {Wilson}, \&
  {Wills}}]{bennert02}
{Bennert}, N., {Falcke}, H., {Schulz}, H., {Wilson}, A.~S., \& {Wills}, B.~J.
  2002, \apjl, 574, L105

\bibitem[{{Bertin} \& {Arnouts}(1996)}]{bertin96}
{Bertin}, E. \& {Arnouts}, S. 1996, \aaps, 117, 393

\bibitem[Carlberg(1990)]{carlberg90}
Carlberg, R.G. 1990, \apj, 350, 505

\bibitem[Croom \etal (2002)]{croom02}
Croom, S.M., Boyle, B.J., Loaring, N.S., Miller, L., Outram, P.J., Shanks, T.,
Smith, R.J. 2002, \mnras, 335, 459

\bibitem[de Vries, Becker, \& White(2003)]{devries03}
de Vries, W.H., Becker, R.H., \& White, R.L. 2003, \aj, 126, 1217 

\bibitem[Gehrels(1986)]{gehrels86}
Gehrels, N. 1986, \apj, 303, 336 

\bibitem[{{Haehnelt} {et~al.}(1998){Haehnelt}, {Natarajan}, \&
  {Rees}}]{haehnelt98}
{Haehnelt}, M.~G., {Natarajan}, P., \& {Rees}, M.~J. 1998, \mnras, 300, 817

\bibitem[{{Haiman} \& {Hui}(2001)}]{haiman01}
{Haiman}, Z.~. \& {Hui}, L. 2001, \apj, 547, 27

\bibitem[Hawkins(2002)]{hawkins02}
Hawkins, M.R.S. 2002, \mnras, 329, 76

\bibitem[Hills(1975)]{hills75}
Hills, J.G. 1975, \nat, 254, 295

\bibitem[Hughes \etal (1992)]{hughes92}
Hughes, P.A., Aller, H.D., \& Aller, M.F. 1992, \apj, 396, 469

\bibitem[Jakobsen \etal (2003)]{jakobsen03}
Jakobsen, P., Jansen, R.A., Wagner, S., \& Reimers, D. 2003 \aap, 397, 891 

\bibitem[{{Kauffmann} \& {Haehnelt}(2000)}]{kauffmann00}
{Kauffmann}, G. \& {Haehnelt}, M. 2000, \mnras, 311, 576

\bibitem[{{Martini}(2003)}]{martini03}
{Martini}, P. 2003, to appear in ``Carnegie Observatories Astrophysics 
Series, Vol. 1: Coevolution of Black Holes and Galaxies,'' ed. L. C. Ho
(Cambridge: Cambridge Univ. Press) (astro-ph/0304009)

\bibitem[{{Martini} \& {Weinberg}(2001)}]{martini01a}
{Martini}, P. \& {Weinberg}, D.~H. 2001, \apj, 547, 12

\bibitem[Richards \etal (2002)]{richards02}
Richards, G.T. \etal\ 2002, \aj, 123, 2945 

\bibitem[Schneider \etal (2002)]{schneider02}
{Schneider}, D. \etal\ 2002, \aj, 123, 567 

\bibitem[Scott \etal (2000)]{scott00}
Scott, J., Bechtold, J., Dobrzycki, A., Kulkarni, V.P. 2000, \apjs, 130, 67

\bibitem[Simonetti \etal (1985)]{simonetti85}
Simonetti, J.H., Corder, J.M., \& Heeschen, D.S. 1985, \apj, 296, 46 

\bibitem[So\l tan(1982)]{soltan82}
So\l tan, A. 1982, \mnras, 200, 115

\bibitem[Tyson \etal (2002)]{tyson02}
Tyson, J. A., \& LSST Collaboration 2002, Proc. SPIE, 4836, 10

\bibitem[Ulmer(1999)]{ulmer99}
Ulmer, A. 1999, \apj, 514, 180

\bibitem[{York} {et~al.}(2000)]{york00}
{York}, D.~G., \etal\ 2000, \aj, 120, 1579

\bibitem[Young, Sheilds, \& Wheeler(1977)]{young77}
Young, P.J., Shields, G.A., Wheeler, J.C. 1977, \apj, 212, 376

\bibitem[{{Yu} \& {Tremaine}(2002)}]{yu02}
{Yu}, Q. \& {Tremaine}, S. 2002, \apj, 335

\end{thebibliography}
\end{document}